\documentclass[aps,prb,twocolumn,notitlepage,amsfonts,citeautoscript,superscriptaddress]{revtex4-2}

\usepackage{amsmath,amssymb,mathrsfs}
\usepackage{latexsym}
\usepackage{graphicx} 

\usepackage{grffile}
\usepackage{graphicx,epstopdf,color}
\usepackage{amsfonts}
\usepackage[dvipsnames]{xcolor}
\usepackage{url}
\usepackage{soul}
\usepackage{cancel}

\usepackage{hyperref}
\hypersetup{
    colorlinks = true,
    linkcolor = MidnightBlue,
    citecolor = MidnightBlue,
    linkbordercolor = {white},
    urlcolor = RoyalBlue
}
\usepackage{cleveref}
\crefname{equation}{Eq.}{Eqs.}
\Crefname{equation}{Equation}{Equations}
\crefname{figure}{Fig.}{Figs.}
\Crefname{figure}{Figure}{Figures}
\crefname{section}{Sec.}{Secs.}
\Crefname{section}{Section}{Sections}
\crefname{appendix}{Appendix}{Apps.}
\Crefname{appendix}{Appendix}{Apps.}
\crefname{paragraph}{Sec.}{Secs.}
\crefname{table}{Table}{Tables}

\usepackage{bm}
\newcommand{\textalert}[1]{}

\newcommand{\ket}[1]{\left|#1\right\rangle}
\newcommand{\bra}[1]{\left\langle#1\right|}

\newcommand{\braket}[2]{\bigl\langle#1\bigl|\bigr.#2\bigr\rangle}

\newcommand{\eps}{\varepsilon}

\newcommand{\GCD}{\mathrm{GCD}}

\newcommand{\ag}[1]{{\color{Orange}{#1}}}
\renewcommand{\ag}[1]{{#1}}
\newcommand{\ab}[1]{{\color{MidnightBlue}{#1}}}

\def\ie{\emph{i.e.},\ }

\usepackage{xr}

\graphicspath{{Figures/}}

\usepackage[caption=false]{subfig}
\usepackage{epstopdf}

\newcommand{\wa}{\omega_a}
\newcommand{\wb}{\omega_b}
\newcommand{\wc}{\omega_c}

\newcommand{\uca}{u_{ca}}
\newcommand{\ucb}{u_{cb}}
\newcommand{\ucc}{u_{cc}}

\newcommand{\Wa}{\omega_a}
\newcommand{\Wb}{\omega_b}
\newcommand{\Wc}{\omega_c}

\newcommand{\ha}{\hat{a}}
\newcommand{\hb}{\hat{b}}
\newcommand{\hc}{\hat{c}}
\newcommand{\hba}{\hat{a}}
\newcommand{\hbb}{\hat{b}}
\newcommand{\hbc}{\hat{c}}

\newcommand{\balb}{\alpha_b}
\newcommand{\balc}{\alpha_c}

\newcommand{\snr}{\text{SNR}}
\newcommand{\tramp}{T_\text{ramp}}
\newcommand{\ns}{~\text{ns}}
\newcommand{\GHz}{~\text{GHz}}

\newcommand{\mcs}{~\text{\textmu s}}

\begin{document}
\title{Engineering, control and longitudinal readout of Floquet qubits}
\author{Anthony Gandon}
\affiliation{Institut Quantique and D\'epartement de Physique, Universit\'e de Sherbrooke, Sherbrooke, Qu\'ebec, J1K 2R1, Canada}
\affiliation{Mines ParisTech, PSL Research University, F-75006 Paris, France}
\author{Camille Le Calonnec}
\affiliation{Institut Quantique and D\'epartement de Physique, Universit\'e de Sherbrooke, Sherbrooke, Qu\'ebec, J1K 2R1, Canada}
\affiliation{University of Strasbourg and CNRS, CESQ and ISIS (UMR 7006), 67000 Strasbourg, France}
\author{Ross Shillito}
\affiliation{Institut Quantique and D\'epartement de Physique, Universit\'e de Sherbrooke, Sherbrooke, Qu\'ebec, J1K 2R1, Canada}
\author{Alexandru Petrescu}
\affiliation{Institut Quantique and D\'epartement de Physique, Universit\'e de Sherbrooke, Sherbrooke, Qu\'ebec, J1K 2R1, Canada}
\author{Alexandre Blais}
\affiliation{Institut Quantique and D\'epartement de Physique, Universit\'e de Sherbrooke, Sherbrooke, Qu\'ebec, J1K 2R1, Canada}
\affiliation{Canadian Institute for Advanced Research, Toronto, M5G1M1 Ontario, Canada}

\begin{abstract}
    Properties of time-periodic Hamiltonians can be exploited to increase the dephasing time of qubits and to design protected one- and two-qubit gates. Recently, Huang \textit{et al.}~[Phys. Rev. Applied {\bf 15}, 034065 (2021)] have shown that Floquet states offer a manifold of working points with dynamical protection larger than the few, usual, static sweet spots. Here, we show how Floquet theory, often used on systems with a single drive tone, can be used to describe approaches to robustly control Floquet qubits in the presence of multiple commensurate drive tones. Using this formulation, we introduce a longitudinal readout protocol to measure the Floquet qubit without the need of first adiabatically mapping the Floquet states back to the static qubit states, resulting in a significant speedup in the measurement time of the Floquet qubit. The analytical approach developed here can be applied to any Hamiltonian involving a small number of  distinct drive tones, typical in the study of standard parametric gates for qubits outside of the rotating-wave approximation.
\end{abstract}

\maketitle
\section{Introduction}
\label{Section_I_Introduction}

Realizing the promises of quantum computing requires the ability to manipulate and measure the state of quantum devices with high fidelity. An outstanding challenge towards reaching this goal is the realization of fast and high-fidelity entangling gates with large on-off ratio. An approach to turn on and off entangling interactions is to frequency tune pairs of qubits in and out of resonance. However, noise in the control parameter allowing to tune the qubit frequency introduces an additional source of dephasing, which can be mitigated by operating the qubits at sweet spots where they are first-order insensitive to this noise channel \citep{vionManipulatingQuantumState2002}.

In superconducting qubits such as the transmon \cite{kochChargeinsensitiveQubitDesign2007a}, tuning the qubit frequency is most commonly accomplished by threading a loop with magnetic flux. While for static dc flux bias there are few sweet spots per single flux period $\Phi_0$, it was recently shown that tailored ac modulation of the flux or direct voltage drive on the qubits extends these few static sweet spots to a larger class of dynamical sweet spots \citep{huangEngineeringDynamicalSweet2020,didierFluxControlSuperconducting2019,guoDephasinginsensitiveQuantumInformation2018,didierACFluxSweet2019,valeryDynamicalSweetSpot2021}. This results in more flexibility in choosing the operating points to both maximize dephasing times and facilitate two-qubit gates. 

In the presence of a continuous ac drive, the static computational basis of the qubits is replaced by a set of eigenstates of the periodic Floquet Hamiltonian, also known as Floquet states~\citep{grifoniDrivenQuantumTunneling1998b,chuFloquetTheoremGeneralized2004b}. Protocols for initialization, readout, single-qubit operations and entangling gates on these Floquet qubits have been theoretically proposed~\citep{huangEngineeringDynamicalSweet2020} and experimentally investigated with improvements in the dephasing time~\citep{mundadaFloquetengineeredEnhancementCoherence2020}. In contrast to static qubits, Floquet qubits can be frequency-tuned in a large frequency range by changing the parameters of the drive. This tunability can be used, for example, to implement single-qubit phase gates, but also to bring together pairs of Floquet qubits to activate SWAP-type interactions~\citep{huangEngineeringDynamicalSweet2020}. On the other hand, for X-type single-qubit gates or for two-qubit gates such as the cross-resonance~\citep{rigettiFullyMicroTunable2010}, a second drive is introduced to induce transitions between the Floquet-qubit states~\cite{huangEngineeringDynamicalSweet2020}. 
Moreover, as shown in Refs.~\cite{huangEngineeringDynamicalSweet2020,mundadaFloquetengineeredEnhancementCoherence2020}, the readout of the Floquet qubit can be performed in a two-step process: the Floquet qubit is first mapped to the laboratory-frame qubit by adiabatically turning off the drive. At that point, a usual dispersive readout is performed by driving a cavity coupled to the qubit \citep{Blais2021}.\\

\ag{In this work, we describe the dynamics of qubits subject to two commensurate drives within Floquet theory. First, a continuous drive on the qubit, which we will refer to as the Floquet drive, gives rise to the set of Floquet modes, a two-dimensional subspace of which encodes the Floquet qubit. Control of this Floquet qubit is achieved with at least a second tone, whose frequency is adjusted to address transitions within the Floquet quasienergy spectrum. The technique presented below allows one to extract gate rates for such single-qubit operations from the quasienergy spectrum corresponding to the system with two (or, more generally, multiple) commensurate drives. The requirement of commensuration between the Floquet drive and the control drive allows us to numerically obtain the Floquet quasienergy spectrum without resorting to less practical approaches, such as many-mode Floquet theory~\citep{hoFloquetLiouvilleSupermatrixApproach1986, shirleySolutionSchrodingerEquation1965a, shillSEMICLASSICALMANYMODEFLOQUET1983}.

Building on these ideas, we then show how longitudinal readout of a Floquet qubit can be achieved with the help of a readout cavity interacting with the qubit via a parametrically driven coupler.} Thanks to both the longitudinal nature of this interaction -- which is known to lead to fast qubit measurements \citep{didierFastQuantumNondemolition2015} -- and to the fact that there is no need to map the Floquet qubit back to the undriven qubit states before readout, we find from numerical simulations that this approach can lead to fast and high-fidelity Floquet qubit readout. A superconducting circuit design for this readout protocol is proposed.

The paper is structured as follows. In \cref{Section_II_Floquet_Framework}, we review the framework of Floquet theory. In \cref{Section_III_Single_qubit_operations} we then recall how the $X$-gate can be implemented on Floquet states by adding a second drive to the qubit, and then use Floquet theory to economically obtain the gate 
rate without simulating its full time dynamics~\cite{Petrescu2021}.
\ag{In \cref{Section_IV_Readout_of_floquet_states} we demonstrate the feasibility of dynamical longitudinal readout of Floquet states using an additional drive and compare our analytical results to full numerics.}
\ag{For the completeness, we finally explore in \cref{Section_V_Initialization_of_arbitrary_states} the procedure for initializing a Floquet qubit in different parameter regimes}


\section{Floquet Framework} 
\label{Section_II_Floquet_Framework}


The driven quantum systems considered here are part of a larger class of systems evolving under a time-periodic Hamiltonian with period $T=2\pi/\omega_{d}$ and which are efficiently described by the Floquet formalism~\citep{grifoniDrivenQuantumTunneling1998b,chuFloquetTheoremGeneralized2004b}. \ag{The total Hamiltonian $H(t)= H_{\text{s}}+V(t)$ includes the time-independent system Hamiltonian $H_{\text{s}}$ in the absence of the external fields and with a finite Hilbert-space dimension $N_s$, and the contribution of the driving fields $V(t)$ which we suppose $T$-periodic in time.}

Based on the symmetry of the full Hamiltonian under time translation $t\rightarrow t+T$, the Floquet theorem states the existence of a \ag{complete} set of $N_s$ solutions to the time-dependent Schr\"odinger equation of the form $\ket{\psi_{n}(t)} = e^{-i\epsilon_nt}\ket{\phi_{n}(t)}$. Here, the \textit{Floquet modes} $\ket{\phi_{n}(t)}$ are $T$-periodic in time, and the \textit{quasienergies} $\epsilon_{n}$ are real-valued coefficients which are invariant under translation by multiples $k$ of the drive frequency $\omega_{d}$.
\ag{Note that these quasienergies take into account the various effects of the drive such as the ac-Stark shift.} The term \textit{quasienergies} refers to representatives of equivalence classes, often chosen in the first Brillouin zone $[-\omega_{d}/2, \omega_{d}/2]$. With appropriately designed driving protocols, one can
continuously map the eigenstates of $H_s$ to the Floquet states of $H(t)$. 

In this context, dynamical protection consists of operating the Floquet qubit at extrema of the quasienergy difference with respect to the drive parameters~\citep{didierACFluxSweet2019, didierFluxControlSuperconducting2019, huangEngineeringDynamicalSweet2020}. As shown by \textcite{huangEngineeringDynamicalSweet2020}, dynamical sweet spots represent manifolds in parameter space, in contrast with the few isolated static sweet spots that are found in the absence of a drive. This allows for an increased freedom in the parameter choice that can be used to operate the Floquet qubit while being protected from low-frequency noise, which translates to high coherence times. This property is compatible with single and two-qubit gate operations, making Floquet qubits promising candidates for quantum information processing.

\ag{Applying logical gates on a Floquet qubit typically requires two distinct time-dependent terms:}
\begin{equation}
\label{General_Floquet_Hamiltonian_Two_Tones}
    H(t) = H_{\text{s}}+V_1(t)+V_2(t),
\end{equation}
with $V_1,V_2$ respectively $2\pi/\omega_{d1}$- and $2\pi/\omega_{d2}$- periodic in time. The Floquet qubit is generated by the Floquet drive $V_1(t)$ together with the system Hamiltonian, while $V_2(t)$ is only switched on during the gate without any \emph{a priori} link between the frequencies $\omega_{d1}$ and $\omega_{d2}$. 
\ag{We will limit the discussion to these two distinct drives and their harmonics, even if the scheme we present is more general. We note that in the rotating-wave approximation (RWA), the two distinct frequencies can result in a single explicit time dependence of the Hamiltonian at the difference frequency $\omega_{d1} - \omega_{d2}$. The Floquet analysis aims at describing the dynamics of driven systems without such simplification, and at efficiently extracting parameters describing the gate Hamiltonian from the Floquet spectrum of the two-tone Hamiltonian of \cref{General_Floquet_Hamiltonian_Two_Tones}.
}

The extension of Floquet theory to commensurate frequencies is straightforward,  using the greatest common divisor frequency $\omega_{\GCD}(\omega_{d1}, \omega_{d2})$, to be defined below, as the new frequency of a single-tone Floquet system for the duration of the gate. This result translates into new Floquet states and quasienergies for \cref{General_Floquet_Hamiltonian_Two_Tones} that can be numerically evaluated. However, the resulting period can be orders of magnitude greater than the timescales $2\pi/\omega_{d1}$ and $2\pi/\omega_{d2}$. In practice, this can lead to long simulation times.


\section{Single-qubit operations} 
\label{Section_III_Single_qubit_operations}

Approaches to realize $X$, $\sqrt{X}$ and single-qubit phase gates on Floquet qubits were proposed in Ref.~\citep{huangEngineeringDynamicalSweet2020}, with fidelities exceeding $99.99\%$ and gate durations on the order of tens of nanoseconds obtained from numerical simulations of the system's time evolution. Here, \ag{we obtain the rate of Rabi oscillations between the quasimodes of the Floquet qubit directly from the Floquet spectrum of \cref{General_Floquet_Hamiltonian_Two_Tones}, without a complete simulation of the dynamics of the system during the $\sqrt{X}$ and $X$ gates. We work with a two-level system (TLS), as Ref. \citep{huangEngineeringDynamicalSweet2020}.
To build intuition in \cref{Section_III_Subsection_I_XGate_in_the_RWA}, we first do an approximate calculation of the most favorable control tone and the associated gate rate for the $X$-gate, resorting to simplifying assumptions such as the RWA on the driven TLS.}
Then, in \cref{Section_III_Subsection_II_Two_tone_floquet_analysis}, we perform a full two-tone Floquet analysis according to the discussion above ; the results of this subsection apply to any Hamiltonian of the form of \cref{General_Floquet_Hamiltonian_Two_Tones}, and without any use of the RWA. 

\subsection{Floquet qubit and X-Gate in the RWA}
\label{Section_III_Subsection_I_XGate_in_the_RWA}

For completeness, we first review the Hamiltonian of a TLS in the presence of a Floquet drive [$V_1(t)$ in \cref{General_Floquet_Hamiltonian_Two_Tones}] which takes the generic form
\begin{equation}
\label{TLS_Floquet_Hamiltonian}
  H(t) = \frac{\omega_0}{2}\sigma_z + \ag{2}\varepsilon_{d1}\cos(\omega_{d1}t)\sigma_x.
\end{equation}
\ag{In a frame rotating at 
$\omega_{d1}$, $H(t)$ takes the form
\begin{equation}
\label{TLS_Floquet_Hamiltonian_Interaction}
  H'(t) = \frac{\Delta}{2}\sigma_z + \varepsilon_{d1}\sigma_x +
  \varepsilon_{d1}(e^{2i\omega_{d1}t}\sigma_+ + e^{-2i\omega_{d1}t}\sigma_-),
\end{equation}
where $\Delta=\omega_0 - \omega_{d1}$. For a low-amplitude and nearly resonant drive, $\omega_{d1}\gg \lbrace \Delta,\varepsilon_{d1}\rbrace$ and so the oscillating terms in \cref{TLS_Floquet_Hamiltonian_Interaction} can be neglected in the RWA. The resulting Hamiltonian can be diagonalized and yields, after a transformation back to the laboratory frame, the Floquet states and quasi-energies of the driven TLS:}
\begin{equation}
\begin{split}
\label{TLS_Floquet_Modes}
    \epsilon_{0,1} &= \pm\sqrt{\left(\frac{\Delta}{2}\right)^2+\varepsilon_{d1}^2}, \\
    \ket{\phi_{0,1}(t)} &= \frac{e^{+i\omega_{d1}t/2}}{\sqrt{\varepsilon_{d1}^2+(\epsilon_{0,1}-\frac{\Delta}{2})^2}}\begin{pmatrix} |\varepsilon_{d1}|e^{-i\omega_{d1}t} \\ \epsilon_{0,1}-\frac{\Delta}{2}\end{pmatrix}.
\end{split}
\end{equation}
In the aforementioned regime, the Floquet modes $\ket{\phi_{0,1}(t)}$ are located near the equatorial plane of the Bloch sphere. 

Following \textcite{huangEngineeringDynamicalSweet2020}, we introduce a control drive in addition to the Floquet drive to realize the $X$-gate. \ag{To induce Rabi oscillations of the Floquet states, this second drive is chosen along the $Z$-axis \ie orthogonal to the Floquet qubit such that the Hamiltonian in the laboratory frame now reads}
\begin{equation}
\label{TLS_FLoquet_with_Drive}
  H(t)= \frac{\omega_0}{2}\sigma_z+\ag{2}\varepsilon_{d1}\cos\left(\omega_{d1} t\right)\sigma_x+\ag{2}\varepsilon_{d2}(t)\cos\left(\omega_{d2} t\right)\sigma_z,\\
\end{equation}
where $\varepsilon_{d2}(t)$ is the amplitude of the second drive and $\omega_{d2}$ its frequency which is chosen to be close to the quasienergy difference 
$\epsilon_{1}-\epsilon_{0}$, according to \cref{TLS_Floquet_Modes} within the RWA. For transmon qubits, a drive along the $Z$ axis can be realized by flux pumping the qubit's SQUID loop~\cite{kochChargeinsensitiveQubitDesign2007a}. Applying the RWA  a second time for the control drive, the effective gate rate on the Floquet qubit takes the form \begin{equation}
\label{Quasienergies_RWA}
    \Omega_{RWA} = 2\sqrt{\left[\frac{\omega_{d2} - (\epsilon_1 - \epsilon_0)}{2}\right]^2+\varepsilon_{d2}^2}.
\end{equation}
Corrections beyond the RWA, applicable also in the more general case of off-resonant drives, can be derived~\cite{mirrahimi2015dynamics}.

\begin{figure}[t]
    \centering
    \includegraphics[width=\columnwidth]{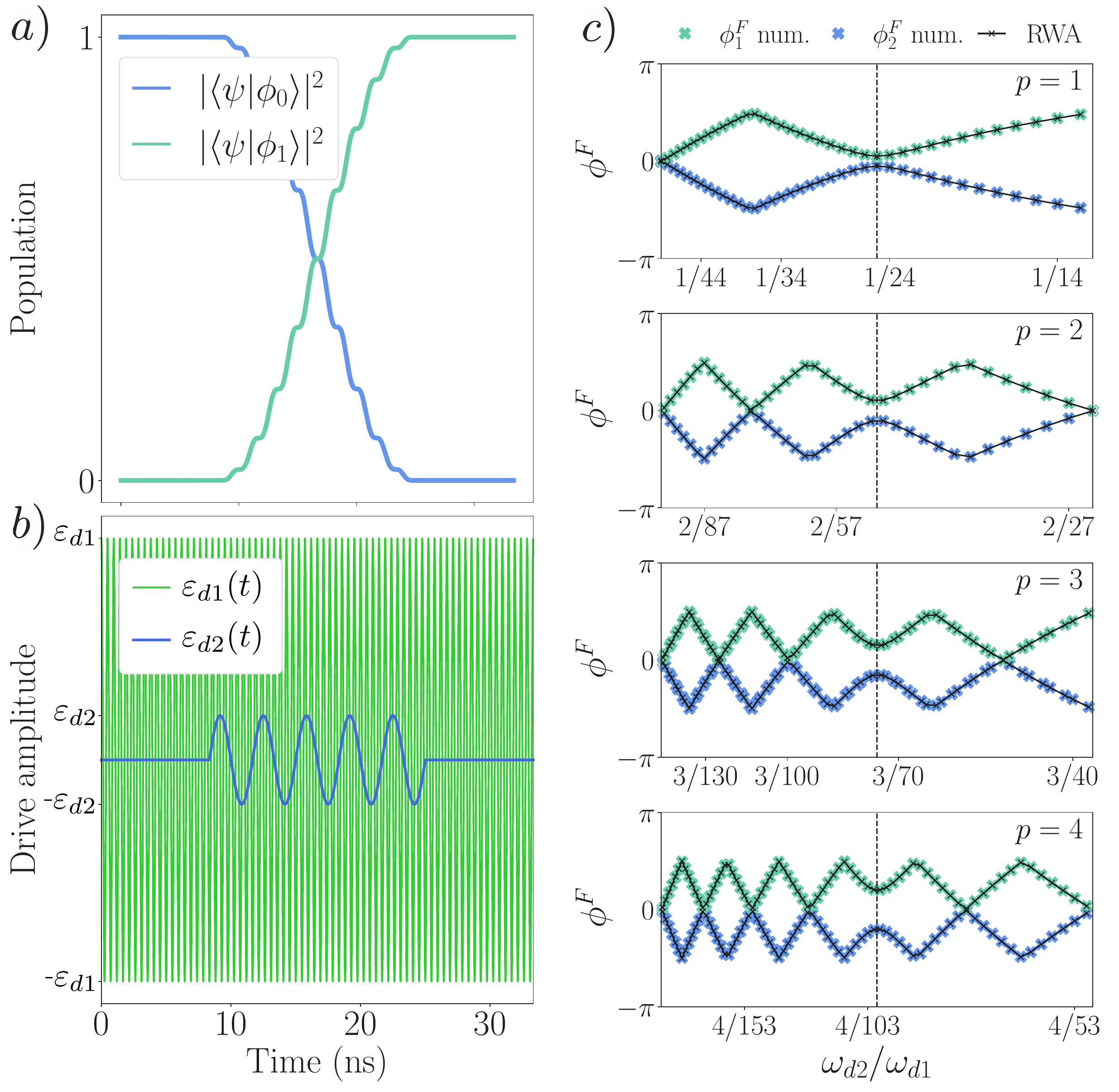}
    \caption{a) Population of the Floquet modes $\ket{\phi_0(t)}$ and $\ket{\phi_1(t)}$ in the numerically evolved state $\ket{\psi(t)}$ as a function of time under the Hamiltonian \cref{TLS_FLoquet_with_Drive} for the parameters found in the main text.
    b) Typical drive amplitude as a function of time. The first drive $\varepsilon_{d1}$ (green line) is used to generate the Floquet qubit states while the second drive $\varepsilon_{d2}$ (blue line) drives Rabi oscillations between these levels.
    c) Four quasiphase spectra for different numerators $p=1, 2, 3, 4$ in the ratio $\omega_{d2}/\omega_{d1}=p/q$. The colored dots are obtained from diagonalization of the propagator associated with \cref{TLS_FLoquet_with_Drive} after a common period $2\pi/\omega_{\GCD}$ and by sweeping the values of $q$. Because of the folded space, several crossings are observed. However, a unique anticrossing corresponding to the resonance of the second drive with the Floquet qubit is observed (vertical line). The full black lines are obtained from  \cref{Quasienergies_RWA}.
    }
    \label{Figure_XGate_Floquet_Qubits}
\end{figure}

As a first illustration of the $X$ gate on the Floquet qubit, we show in \cref{Figure_XGate_Floquet_Qubits}(a) the population of the Floquet modes $\ket{\phi_{0}}$ (blue) and $\ket{\phi_{1}}$ (green) obtained from numerical integration of the Schr\"odinger equation under the Hamiltonian of \cref{TLS_FLoquet_with_Drive} using Qutip~\cite{johanssonQuTiPPythonFramework2013}. 
\ag{The parameters are $\omega_0/2\pi = 5.01\GHz$, $\omega_1/2\pi = 5.00\GHz$, $\omega_2/2\pi = 0.225\GHz$, $\varepsilon_{d1}/2\pi = 0.1\GHz$, $\varepsilon_{d2}/2\pi = 0.03\GHz$.
As illustrated by the blue line in \cref{Figure_XGate_Floquet_Qubits}(b), for simplicity the amplitude of the control drive takes the form of a step function: $\varepsilon_{d2}(t) = \varepsilon_{d2}$ if $T_1<t<T_2$ and 0 else with the
condition that the control tone vanishes at the endpoints, $\cos(\omega_{d2}T_{1,2})=0$.} This tone is switched on for the duration of the gate and, following \cref{TLS_Floquet_Modes}, its frequency $\omega_{d2}$ is chosen to be close to twice the amplitude $\varepsilon_{d1}$ of the Floquet drive, corresponding to the limit $\Delta \ll \epsilon_{d1}$. The green line corresponds to the amplitude of the Floquet drive of amplitude $\varepsilon_{d1}$ and frequency $\omega_{d1}$. With these parameters, the $X$-gate is completed in $\sim 20\ns$ with fidelity 99.99\%. In the following subsection we rely on the exact Floquet two-tone numerical method to obtain the gate rate.

\subsection{Two-tone Floquet analysis}
\label{Section_III_Subsection_II_Two_tone_floquet_analysis}

We now turn to an alternative approach to analyze gates on the Floquet qubit based on a two-tone Floquet analysis. First, the Floquet spectrum is obtained from the Hamiltonian of \cref{TLS_FLoquet_with_Drive} with respect to the second drive frequency $\omega_{d2}$ without requiring the RWA.
To obtain this spectrum, we regroup the time-dependent terms into a single quasi-periodic drive $V(t) = V_1(t) + V_2(t)$. In our case of commensurate frequencies $\omega_{d1}$ and $\omega_{d2}$, then the periodicity of $V(t)$ is given by the greatest common divisor frequency $\omega_{\GCD}$.
The quasienergy spectrum is probed by sweeping the control frequency $\omega_{d2}$ while keeping the Floquet drive frequency $\omega_{d1}$ fixed.
However, because the quasienergies are only defined modulo $\omega_{\GCD}$, and because this quantity will strongly depend on the chosen $\omega_{d2}$, it is not possible to define a continuous quasienergy spectrum. 
However, when the drive frequencies can be written as an irreducible fraction $\omega_{d1}/\omega_{d2} = p/q$, the frequency $\omega_{\GCD}$ can be expressed as $\omega_{\GCD} = \omega_{d1}/p = \omega_{d2}/q$. Here, $\omega_{d1}$ is taken as a fixed parameter such that 
each numerator $p$ corresponds to a distinct first Brillouin zone. For each numerator $p$, we can then introduce a discrete quasienergy spectrum satisfying $\omega_{d2}=\omega_{d1}\times q/p$ for $q \ge 0$ integer. \ag{To compare Brillouin zones of different sizes, we normalize the Floquet quasienergy spectrum $\epsilon_{0,1}(\omega_{d2})$ defined over the 
first Brillouin zone $[-\omega_{\GCD}/2,\omega_{\GCD}/2]$ to obtain the Floquet quasiphase spectrum defined as $\phi^F_{0,1}(\omega_{d2}) = \epsilon_{0,1}(\omega_{d2})\times 2\pi/\omega_{\GCD}$ over $[-\pi, \pi]$.}

In \cref{Figure_XGate_Floquet_Qubits}(c), we plot the quasiphase spectra associated with the Hamiltonian \cref{TLS_FLoquet_with_Drive} for commensurate ratios $\omega_{d2}/\omega_{d1}$ with small numerators. The different subplots illustrate the discrete quasiphase spectra for different values of the numerator $p$. \ag{The continuous black curves are obtained from the RWA form of the gate rate in  \cref{Quasienergies_RWA}. It should be seen as continuous modulo~$2\pi$ but at this stage it is hard to unfold, which also explains why we distinguish the computed eigenvalues by sorting them. The visual horizontal contraction of the curve for increasing $p$ is the direct effect of the normalisation of smaller first Brillouin zones with an increasing number of foldings.} The difference between the two quasiphases $\phi^F_{0}$ and $\phi^F_{1}$ exhibits a local minimum over all the subplots around $\omega_{d2}/\omega_{d1}\approx 0.04 \approx 2\varepsilon_{d1} / \omega_{d1}$ which is linked to an avoided crossing characterizing the resonance of the control drive with the Floquet qubit and thus yields the gate rate (see \cref{Appendix_II_Numerical_approach} for details on obtaining these plots). The analytical approximation of the previous subsection is in good agreement when the parameter choice satisfies the RWA conditions, as is the case here. 

A precise numerical estimate of the size of this anticrossing, valid without any RWA, is obtained by increasing the maximum allowed numerator at the cost of longer simulations. In particular, \cref{Figure_Appendix_A_Wall_time}(b) in \cref{Appendix_II_Numerical_approach} shows that a lower bound of around $10^{-3}~$rad for the precision which can be obtained within the RWA even with this favorable parameter choice.
Here, we avoid intensive numerical simulations by taking advantage of Dysolve \citep{shillitoFastDifferentiableSimulation2020a}, a semi-analytical solver whose performance in determining the time-evolution operator of the Hamiltonian in \cref{TLS_FLoquet_with_Drive} is discussed in \cref{Appendix_II_Numerical_approach}.
The validity of the approach presented here goes beyond the two-level approximation used in this work and can, for example, easily be extended to multi-level Floquet qubits.


\section{Longitudinal Floquet qubit readout}
\label{Section_IV_Readout_of_floquet_states}

\ag{Having described an approach based on the quasiphase to precisely estimate gate rate for Floquet qubits, we now turn to readout of those qubits. In Refs.~\cite{huangEngineeringDynamicalSweet2020,mundadaFloquetengineeredEnhancementCoherence2020,Deng2015}, this is realized by adiabatically mapping the Floquet logical states back to the original undriven qubit states, followed by a usual dispersive qubit readout \cite{Blais2021}. This two-step process leads to a longer measurement time than strictly necessary. Here, we introduce an approach to engineer a longitudinal coupling between the Floquet qubit and a readout mode by using a modulated transverse coupling on the qubit.} 
Because of its longitudinal nature, the approach we introduce here can reach a large signal-to-noise ratio (SNR) in a measurement time that is small compared to the usual dispersive readout of circuit QED \cite{didierFastQuantumNondemolition2015}. This approach bears similarities with the stroboscopic measurements of Ref.~\citep{eddinsStroboscopicQubitMeasurement2018} and the Kerr-cat qubit readout of Ref.~\citep{grimmKerrCatQubitStabilization2020}.

\subsection{Engineered longitudinal coupling}
\label{Section_IV_Subsection_I_Derivation_of_the_readout}

Our longitudinal readout is based on coupling between a Floquet qubit and a readout cavity. In the laboratory frame, the Hamiltonian reads
\begin{equation}
    \label{TLS_Readout_Theorical_Hamiltonian}
    H(t) =
    \frac{\omega_0}{2}\sigma_z+\ag{2}\varepsilon_{d1}\cos(\omega_{d1}t)\sigma_x+\omega_r\hat{a}^\dag\hat{a}+g(t)(\hat{a}+\hat{a}^\dag)\sigma_x.
\end{equation}
The first two terms are as in \cref{TLS_Floquet_Hamiltonian} and define the Floquet qubit. The last two terms correspond to the free Hamiltonian of a cavity of frequency $\omega_r$ and annihilation operator $\hat a$ coupled to the qubit with a amplitude $g(t)$ which we allow to be time-dependent. In the regime where the detuning $\Delta = \omega_0 - \omega_{d1}$ between the drive and the qubit is small compared to the drive amplitude $\varepsilon_{d1}$, the laboratory frame Pauli operator $\sigma_x$ acts as $\sigma_z$ on the Floquet qubit. As a result, the last term of \cref{TLS_Readout_Theorical_Hamiltonian} results in longitudinal coupling of the Floquet qubit to the cavity mode. In short, the Hamiltonian of \cref{TLS_Readout_Theorical_Hamiltonian} leads to a longitudinal readout of the Floquet qubit, without the need to map back that qubit to the laboratory frame qubit.

To make this more apparent, we move to the interaction frame defined by the time-evolution operator corresponding to the decoupled readout cavity and Floquet qubit
\begin{equation}
\label{TLS_Readout_Propagator}
   U(t,0) = e^{-i\omega_r t \hat{a}^\dag\hat{a} } 
   \sum_{j\in\{0,1\}} \ket{\phi_j(t)}\bra{\phi_j(0)}e^{-i\epsilon_j t}.
\end{equation}
\noindent
In the limit $\Delta/\varepsilon_{d1} \ll 1$ already used in \cref{Section_III_Subsection_I_XGate_in_the_RWA},   \ag{where the Floquet modes \cref{TLS_Floquet_Modes} are nearly eigenstates of $\sigma_x$,} the interaction-picture Hamiltonian takes the form
\begin{equation}
    \label{TLS_Readout_Interaction_Picture}
        H'(t) \approx g(t)\cos(\omega_{0}t)(\hat{a}e^{i\omega_rt}+\hat{a}^\dag e^{-i\omega_rt})\sigma_z^{F}(0).
\end{equation}
Here, we have introduced the interaction-picture Pauli matrices $\sigma_{x,y,z}^{F}(0)$ acting on the basis of the Floquet qubit states
$\left\{\ket{\phi_0(t)},\ket{\phi_1(t)}\right\}$. \ag{Here, the time $t=0$ in the Pauli matrices $\sigma_{z}^{F}(0)$ of \cref{TLS_Readout_Interaction_Picture} refers to the start of the measurement.} Choosing the time-dependent coupling to be of the form $g(t) = \tilde g \cos(\omega_m t)$ with a modulation frequency $\omega_m = \omega_r-\omega_0$ (or, equivalently, $\omega_r+\omega_0$) yields the longitudinal coupling Hamiltonian \cite{didierFastQuantumNondemolition2015}
\begin{equation}
\label{TLS_Readout_Interaction_Picture_Modulated}
    H'(t) \approx \frac{\tilde{g}}{2} (\hat{a}+\hat{a}^\dag)\sigma_z^{F}(0).
\end{equation}
As discussed in Ref.~\cite{didierFastQuantumNondemolition2015}, evolution under this Hamiltonian leads to an optimal separation of the cavity pointer states where the initial cavity vacuum state is displaced 180 degrees out of phase depending on the state of the qubit.

\begin{figure}[t!]
    \centering
    \includegraphics[width=\columnwidth]{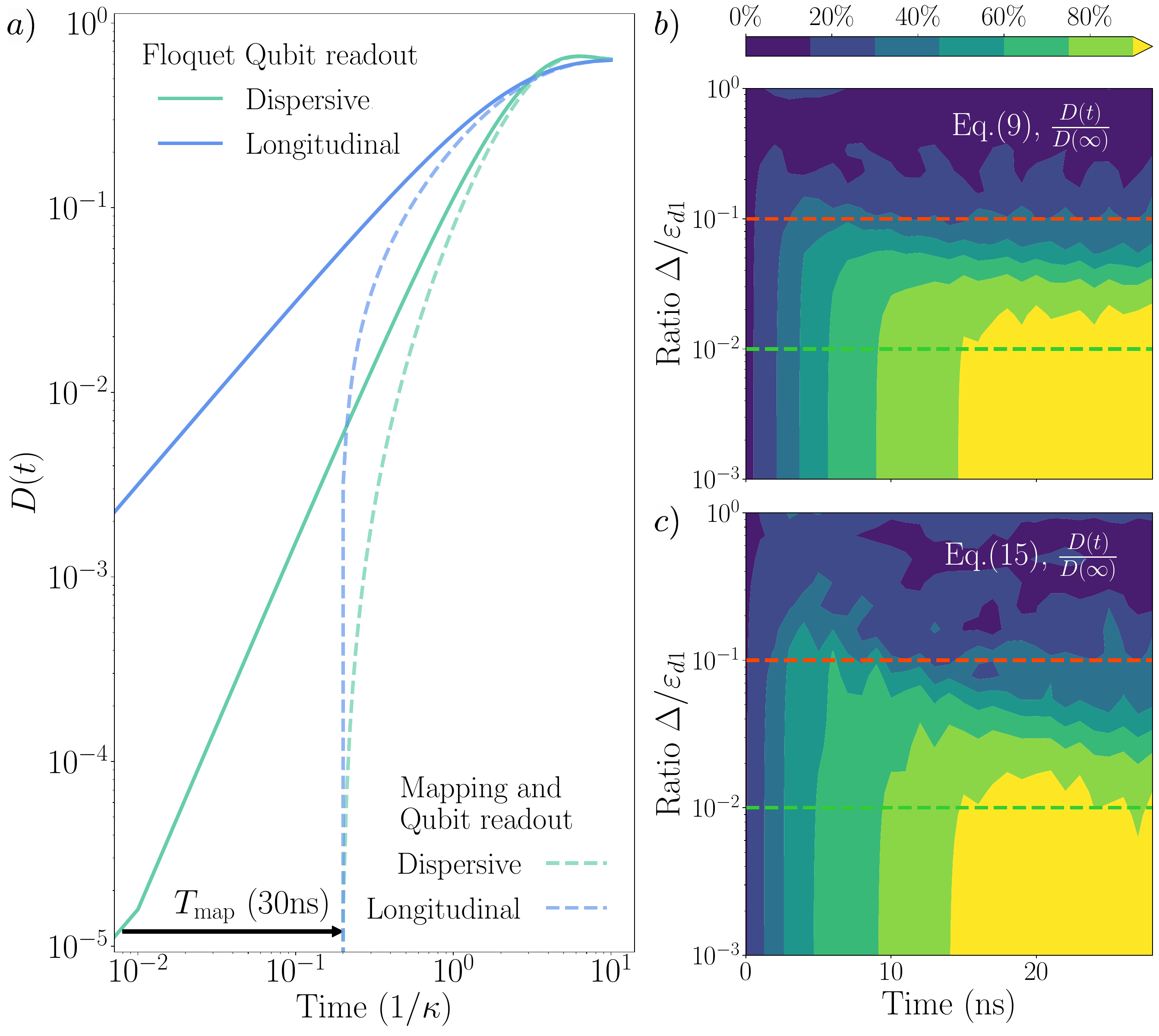}
    \caption{a) Pointer-state separation $D(t)$ as a function of time for longitudinal Floquet readout (full blue line), dispersive readout without state mapping (full green line), dispersive readout with the necessary state mapping with a ramp time of $T_\mathrm{map}=30\ns$ (dashed green line). The dashed blue line corresponds to the hypothetical situation where longitudinal readout is preceded by a mapping stage. This information is included only for comparison. b) Ratio of the pointer state separation $D(t)$ and steady-state separation \ag{$D(\infty) = 0.47 \approx \tilde g /\kappa = 0.5 $} as obtained from numerical integration under \cref{TLS_Readout_Theorical_Hamiltonian} as a function of time and for different ratios $\Delta/\varepsilon_{d1}$. As expected, for small $\Delta/\varepsilon_{d1}$ the pointer states follow the ideal longitudinal dynamics expected from \cref{TLS_Readout_Interaction_Picture_Modulated}. c) Pointer state separation $D(t)$ over the steady-state separation $D(\infty)$ as found from numerical simulation of the system dynamics under the Hamiltonian of \cref{Eq:3KNO}. \ag{Here, $D(\infty)= 0.55$ is numerically evaluated at long times and at small $\Delta/\varepsilon_{d1}$, corresponding to the average value of the bottom-right corner in panel (c). }
    As in panel (b), the pointer state dynamics follow the expected behavior for small with $\Delta/\varepsilon_{d1}$. The parameters used in panel (c) are $\{ \omega_a, \omega_b, \omega_c\}/2\pi = \{8.2,  5.2,7.78\}\GHz$, $\{\balb/2,\balc/2\}/2\pi =\{-0.17,0.4\}\GHz$, $\{g_a,g_b\}/2\pi = \{0.2,0.2\}\GHz$, \ag{$\tilde{\epsilon}_{d1}/2\pi = 0.35\GHz$}, and $\kappa/2\pi = 0.05\GHz$.
    }
    \label{Figure_Pointer_state_separation}
\end{figure}

To compare this Floquet longitudinal readout to the approach based on an adiabatic map followed by a dispersive readout of Refs.~\cite{huangEngineeringDynamicalSweet2020,mundadaFloquetengineeredEnhancementCoherence2020,Deng2015}, we present in \cref{Figure_Pointer_state_separation}(a) the main characteristics of these measurements and highlight their differences. There, we plot the pointer state separation $D(t)=\left|\langle \hat{a}\rangle_0(t) -\langle \hat{a}\rangle_1(t)\right|$\ag{, as obtained from numerical integration over a measurement time $T$ of the Lindblad master equation \cite{Blais2021}
\begin{equation}
    \dot{\hat{\rho}} = -i[\hat{H}(t),\hat{\rho}] + \kappa  \mathcal{D}[\hat{a}]\hat{\rho},
    \label{eqn:MasterEquation}
\end{equation}
under the Hamiltonian of \cref{TLS_Readout_Theorical_Hamiltonian} and with the cavity loss at a rate $\kappa$ represented by the dissipator
\begin{equation}
    \mathcal{D}[\hat O]\bullet = \hat O \bullet \hat O^\dag - \frac{1}{2} \left\{\hat O^\dag \hat O, \bullet \right\}.
\end{equation}}
The pointer state separation is a helpful proxy for the signal-to-noise ratio (SNR) which, assuming a unit measurement-chain efficiency, can be expressed as \citep{bultinkGeneralMethodExtracting2018}
\begin{equation}
    \label{General_SNR_optimal}
         \snr(T) = \sqrt{2\kappa\int_0^TD(t)^2 dt }.
\end{equation}
For longitudinal coupling, the pointer separation takes the simple form \cite{didierFastQuantumNondemolition2015}
\begin{equation}
    \label{TLS_readout_Pointer_state_displacement}
    D(t) = \frac{\tilde{g}}{\kappa} \left( 1 - e^{-\kappa t/2}\right).
\end{equation}

First ignoring the mapping stage which is necessary for the dispersive readout of Refs.~\cite{huangEngineeringDynamicalSweet2020,mundadaFloquetengineeredEnhancementCoherence2020,Deng2015}, the blue lines in \cref{Figure_Pointer_state_separation} correspond to longitudinal readout while the green lines to dispersive readout for which an expression equivalent to \cref{TLS_readout_Pointer_state_displacement} can be obtained \cite{didierFastQuantumNondemolition2015}. Comparing first the full blue and full green lines, we see that longitudinal readout leads to much faster separation of the pointer states than dispersive readout, even when ignoring the adiabatic mapping stage. When accounting for a mapping time of $T_\mathrm{map}=30\ns$ as in Ref.~\cite{huangEngineeringDynamicalSweet2020} (dashed green line), the advantage of the longitudinal approach over dispersive readout becomes even clearer. 
Finally, as a reference, the dashed blue line corresponds to a situation where the mapping stage is followed by a longitudinal readout. Although this would still lead to a faster separation of the pointer state at short times as compared to the dispersive readout, this illustrates that the main gain in the longitudinal Floquet readout introduced here comes from the fact that mapping to the laboratory frame qubit is not required.

As a further verification, \cref{Figure_Pointer_state_separation}(b) shows the pointer state separation $D(t)$ as obtained from numerical integration of the system dynamics under the laboratory-frame Hamiltonian of \cref{TLS_Readout_Theorical_Hamiltonian} as a function of time and for different ratios $\Delta/\varepsilon_{d1}$. In the laboratory frame, we take the modulated coupling to be of the form $g(t) = \tilde{g} [\cos(\omega_r t-\omega_0 t)+\cos(\omega_r t + \omega_0 t)]$. In each simulation, the initial state of the cavity is chosen to be vacuum and the Floquet qubit is initialized to either $\ket{\phi_0(0)}$ or $\ket{\phi_1(0)}$. For ratios $\Delta/\varepsilon_{d1}<0.01$ (horizontal green dashed line), we find the expected exponential increase up to the steady-states $D(\infty)$ in agreement with the analytical result of \cref{TLS_readout_Pointer_state_displacement} already shown in panel (a). On the other hand, when the Floquet qubit is too far away from resonance $\Delta/\varepsilon_{d1}>0.1$, \ag{the modulated term in \cref{TLS_Readout_Theorical_Hamiltonian} is no longer purely longitudinal for the Floquet qubit. As a result,} the separation between the pointer states does not follow the trajectory predicted by \cref{TLS_readout_Pointer_state_displacement} and the readout is suboptimal (horizontal red dashed line).

\subsection{Superconducting circuit implementation}
\label{Section_IV_Subsection_III_Circuit_implementation}

\begin{figure}[t!]
    \centering
    \includegraphics[width=0.8\columnwidth]{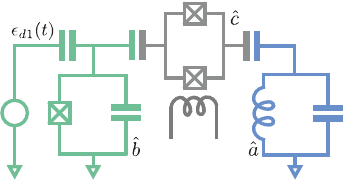}
    \caption{Possible realization of the longitudinal Floquet readout. A driven transmon qubit (green) is coupled to a readout cavity (blue) via a flux modulated coupler (gray).}
    \label{Figure_Appendix_B_CQED}
\end{figure}

A possible realization of this longitudinal Floquet readout with superconducting quantum circuits is illustrated in \cref{Figure_Appendix_B_CQED}. Here, a transmon qubit ($\hat b$) interacts with a readout cavity ($\hat a$) via a flux-tunable coupler ($\hat c$). This system can be modeled as a triplet of coupled Kerr oscillators \cite{Petrescu2021}
\begin{align}
    H = H_a + H_b + H_c(t) + H_g + H_d(t), \label{Eq:3KNO}
\end{align}
where $H_a = \omega_a \hba^\dagger \hba$ corresponds to the linear readout resonator and $H_b = \omega_b \hbb^\dagger \hbb + (\balb/2) \hbb^{\dagger 2} \hbb^2$ to the transmon-like qubit with negative anharmonicity $\balb$. The coupler Hamiltonian takes the same form $H_c = \omega_c(t) \hbc^\dagger \hbc + (\balc/2) \hbc^{\dagger 2} \hbc^2$, except that it is parametrically modulated with $\Wc(t)=\Wc + \delta \Wc (t)$ using a time-dependent flux. The capacitive interactions are modeled by a linear off-diagonal Hamiltonian coupling the bare modes $H_g = g_{ab} \hba^\dagger \hbb + g_{bc} \hbb^\dagger \hbc + g_{ca} \hbc^\dagger \hba + \text{H.c.}$ As shown in \cref{Appendix_IV_Coupler_mediated_interaction}, switching to a normal-mode representation, we can eliminate these bilinear terms to obtain the desired modulated coupling $g(t) = \tilde g \cos(\omega_m t)$ of \cref{TLS_Readout_Theorical_Hamiltonian} between the normal modes corresponding to the qubit and the readout resonator. This is achieved by modulating the coupler frequency at one or both of the sidebands $\omega_a \pm \omega_b$. Finally, the drive on the qubit takes the usual form $H_d(t) = -i \eps_{d1}(t) (\hbb- \hbb^\dagger)$.

The coupling strength $\tilde{g}$ depends on the three capacitive couplings, on the placement of the coupler frequency and on the amplitude of the modulation. Here, we choose this frequency to satisfy the constraint $\Wa < \Wc < \Wb$ to avoid excessive asymmetry. \Cref{Figure_Pointer_state_separation}(c) shows the pointer state separation under the evolution generated by the Lindblad master equation corresponding now to the Hamiltonian of \cref{Eq:3KNO}, and dissipation with similar conditions and parameters to those used in \cref{Figure_Pointer_state_separation}(b). At $\Delta/\varepsilon_{d1}$ small, we verify that the cavity pointer state displacement induced in the cavity by the readout of the Floquet States in the simulation of the full system \cref{Eq:3KNO} matches that of the idealizes Hamiltonian \cref{TLS_Readout_Theorical_Hamiltonian}. Importantly, the fast separation of the pointer states at short time is clearly observed.

\section{Initialization of arbitrary Floquet states}
\label{Section_V_Initialization_of_arbitrary_states}

\ag{In \cref{Section_IV_Readout_of_floquet_states}, we showed how to speed up Floquet qubit readout using an effective longitudinal coupling.
In this section, we consider the timescales needed to initialize a Floquet qubit with high fidelity, addressing a shortcut we took in \cref{Section_III_Single_qubit_operations} when making the assumption that the logical Floquet states $\ket{\phi_{0,1}(0)}$ could be efficiently prepared. In particular, we will see that adiabatic state transfer protocols of Refs.~\citep{mundadaFloquetengineeredEnhancementCoherence2020, desbuquoisControllingFloquetState2017a} where the Floquet drive is slowly turned on are not optimal in the small-detuning regime that is advantageous for longitudinal Floquet readout. Instead, we propose an instantaneous ramping protocol which leads to high-fidelity state preparation in that regime.}

\subsection{Adiabatic initialization}
\label{Section_V_Subsection_I_Adiabatic_regime}

\begin{figure}[t!]
    \centering
    \includegraphics[width=\columnwidth]{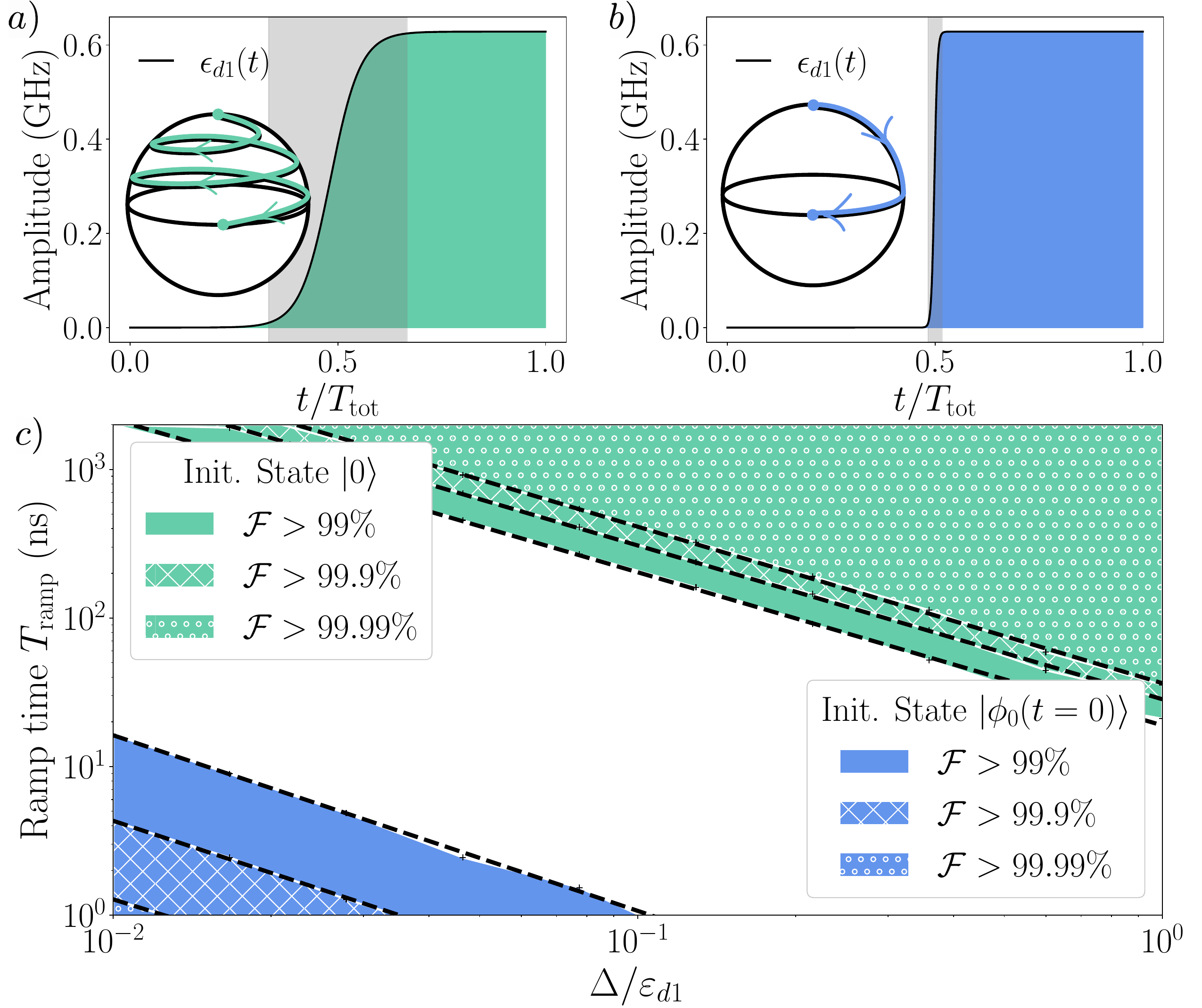}
    \caption{
        Ramp profiles for a) adiabatic and b) instantaneous preparation pulses, as well as illustrative paths on the Bloch sphere. \ag{To facilitate the comparison between panels a) and b), both pulses are illustrated over the total time $T_\mathrm{tot}$ of the adiabatic pulse, with padding added before and after the sudden pulse.}
        c)~Initialization fidelity versus the ratio $\Delta/\varepsilon_{d1}$ and ramp time $T_\mathrm{ramp}$. The different areas correspond to sectors where an initialization fidelity higher than 99\% (plain), 99.9\% (hatched) and 99.99\% (dotted) can be obtained in the adiabatic limit (green) and the instantaneous (blue) regimes.
        }
    \label{Figure_Initialization_Floquet_Qubits}
\end{figure}

\ag{We first consider the timescale needed to adiabatically initialize a Floquet qubit $\ket{\phi_0(t)}$ with a given fidelity. More precisely, we take the system to start in the laboratory frame state $\ket{0}$, and evaluate the fidelity $\mathcal{F} = |\braket{\phi_0(T_\mathrm{tot})}{\psi(T_\mathrm{tot})}|^2$ of the Floquet-state preparation protocol after a time $T_\mathrm{tot}$ by projecting on the desired Floquet state $\ket{\phi_0(T_\mathrm{tot})}$.  \Cref{Figure_Initialization_Floquet_Qubits}(a) illustrates the sigmoid ramp up $\varepsilon_{d1}(t) = {\varepsilon_{d1}}/(1+e^{-2\sigma(t/T_\mathrm{ramp}-1.5)})$ with characteristic time $T_\mathrm{ramp}$ and width $\sigma=4$ used for the Floquet drive amplitude in this adiabatic protocol. The states $\ket{\psi(T_\mathrm{tot})}$ is obtained from numerical simulations of \cref{TLS_Floquet_Hamiltonian} under this drive. Reducing $\tramp$ is expected to worsen the state preparation fidelity of this adiabatic protocol.}

\ag{The preparation fidelity is numerically computed as a function of the ramp time $\tramp$ and for various ratios $\Delta/\varepsilon_{d1}$ of the drive profile, see \cref{Figure_Initialization_Floquet_Qubits}(c). There we}
identify the minimal $\tramp$ corresponding to a fidelity $\mathcal{F}$ larger than 99\% (plain green), 99.9\% (hatched green) and 99.99\% (dotted green) for each ratio $\Delta/\varepsilon_{d1}$. We characterize the boundary of these empirical regions (dashed lines in log-log scale) by fitting an empirical law
\begin{equation}
\label{Initialization_Empirical_Law}
\tramp \times \left|\frac{\Delta}{\varepsilon_{d}}\right|\geq \tau_1,
\end{equation}
where we find $\tau_1=18.9\ns$ for a 99\% fidelity, $\tau_1=28.4\ns$ for 99.9\%, and $\tau_1=36.4\ns$ for 99.99\%. Extrapolating this proportionality relation closer to resonance $\Delta=0$, we obtain the divergence of the adiabatic ramping time already observed in the context of driven two-body quantum systems~\citep{desbuquoisControllingFloquetState2017a}. 
In particular, for the small $\Delta/\varepsilon_{d1}$ used in the longitudinal readout of the previous section, we find that initialization cannot be obtained with times smaller than the adiabatic lower bound $\tramp \approx \tau_1/0.01 = 1.9\mcs$ for a 99\% fidelity and $2.8\mcs$ for a 99.9\% fidelity.

\subsection{Sudden initialization}
\label{Section_V_Subsection_II_Instantaneous_regime}

Because of the long preparation time required with small $\Delta/\varepsilon_{d1}$ which is optimal for the longitudinal readout of \cref{Section_IV_Readout_of_floquet_states}, we now consider an alternative in the form of an instantaneous ramping protocol. Here, the idea consists in first preparing an initial superposition $\alpha\ket{0} + \beta\ket{1}$ of the laboratory frame qubit using standard pulse techniques such as to equal the instantaneous eigenstate $\ket{\phi_0(0)}$ of the desired time-dependent Hamiltonian. An abrupt increase of the drive amplitude $\varepsilon_{d1}(t)$, as illustrated in \cref{Figure_Initialization_Floquet_Qubits}(b), then connects the eigenstate $\ket{\phi_0(0)}$ of the instantaneous Hamiltonian $H(0)$ and the Floquet states $\ket{\phi_0(t)}$ of the time-dependent Hamiltonian $H(t)$ leading to a high-fidelity state preparation.

Computing the fidelity of this protocol as a function of the ramp time and ratio $\Delta/\varepsilon_{d1}$, we find in \cref{Figure_Initialization_Floquet_Qubits}(c) that the high-fidelity region (plain blue) is now delimited in parameter space by an upper bound in log-log scale rather than by a lower bound as was the case for the adiabatic protocol. This upper bound can be expressed as
\begin{equation}
\label{Initialization_Empirical_Law_Instantaneous}
\tramp\times \frac{\Delta}{\varepsilon_{d1}}\leq \tau_2,
\end{equation}
where $\tau_2=0.18\ns$ for a 99\% fidelity, $\tau_2=0.06\ns$ for 99.9\%, and $\tau_2=0.03\ns$ for 99.99\%. Notably, for the ratio $\Delta/\varepsilon_{d1}=0.01$ which led to a fast longitudinal readout, this corresponds to a ramp time as fast as
$18\ns$ (resp.~$6\ns$) to reach 99\% (resp.~99.9\%) fidelity. \ag{In short, the sudden approach to initialization can be used to prepare Floquet qubits with high fidelity in a fraction of the time needed for the adiabatic protocol.} 

\section{Summary}

With the objective of identifying optimal gate parameters for Floquet qubits, we have shown how to define the quasiphase spectra of a static system with two distinct drives and how to extract gate parameters from such spectra. To compensate for the computational cost of this approach, we use the semi-analytic Dysolve method for the integration of the unitary dynamics in our system \cite{shillitoFastDifferentiableSimulation2020a}. In this way, we find a tenfold improvement in simulation time as compared to the QuTiP solver \cite{johanssonQuTiPPythonFramework2013}, opening up a path toward precise quasiphase spectra of complex quantum systems with two drives and a larger Hilbert space. Additionally, we introduce longitudinal Floquet qubit readout which, in contrast with previous methods, does not require mapping the Floquet qubit to the laboratory-frame qubit before the measurement. \ag{Finally, we show how Floquet qubits can be initialized in short times with high preparation fidelity.} 
\ag{Combined with existing procedures for single-qubit gates and two-qubit gates, these results show that quantum information processing can be performed with Floquet qubits without having to move to the underlying static undriven basis which does not benefit from the presence of dynamical sweet spots.}
These results open new possibilities to further optimize logical gates and operations on Floquet qubits using the analytical understanding of the extended Floquet theory when only a few uncorrelated driving frequencies are involved. In future work, we will apply this framework to two-Floquet qubit gates with the objective of identifying optimal gate parameters with an approach that is free of approximations.

\section*{Acknowledgments}
We thank Marie Lu, Jean-Loup Ville, Joachim Cohen, and Ziwen Huang, Jens Koch for stimulating discussions. This work was undertaken thanks to funding from NSERC, the Canada First Research Excellence Fund and the U.S. Army Research Office Grant No.~W911NF-18-1-0411. This material is based upon work supported by the U.S. Department of Energy, Office of Science, National Quantum Information Science Research Centers, Quantum Systems Accelerator.

\bibliography{Bibliography/bib_anthony_05_24_bibtex}

\appendix


\section{Numerical approach for the Two-tone Floquet study}
\label{Appendix_II_Numerical_approach}

\begin{figure}
    \centering
    \includegraphics[width=\linewidth]{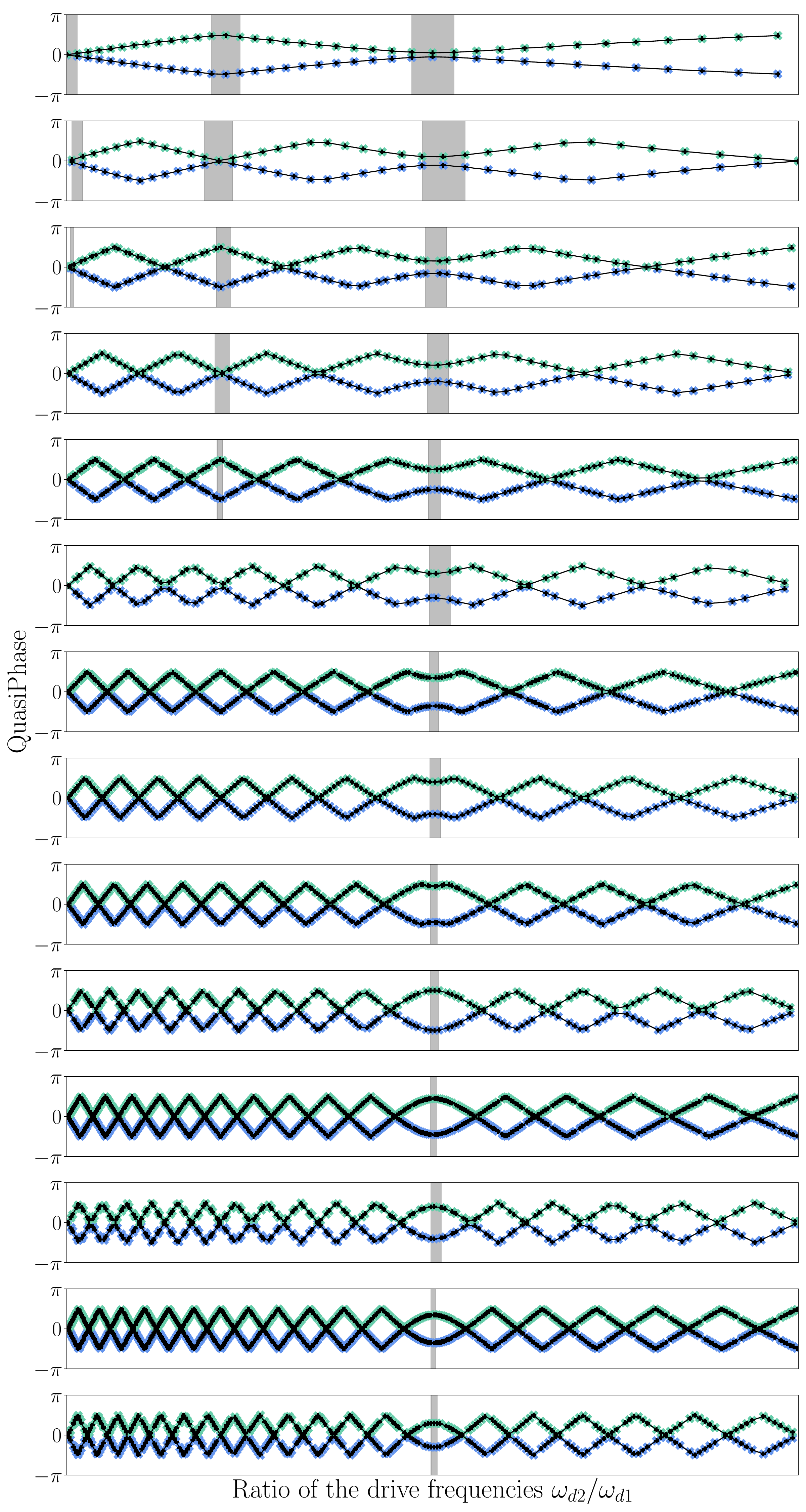}
    \caption{Quasiphase spectrum for numerators $1\leq p_k\leq 15$. The local minima of the avoided crossing can be observed on the first subplots and is characterized by a size $p\times \varepsilon_{d2}$. As this quantity goes beyond $\pi/2$, the local minimum becomes a local maximum in the folded space. The grey areas are used to visualize the triplet  $\left\{\phi_{p_k}[q'-1], \phi_{p_k}[q'], \phi_{p_k}[q'+1]\right\}$ corresponding to local extrema with a non-empty intersection to the grey areas in the above subplots. }
    \label{Figure_Appendix_A_MMFT_Spectrum}
\end{figure}

To obtain the quasiphase spectra in \cref{Figure_XGate_Floquet_Qubits}(c), we build the propagator associated with \cref{TLS_FLoquet_with_Drive} for each ratio $\omega_{d1}/\omega_{d2}=p/q$. For a fixed numerator $p$, we sweep the denominator $q$ and simulate the time dynamics over the period $2\pi p/\omega_{d1}$. For a TLS, diagonalization of the propagator then yields two eigenvalues on the unit circle, which correspond to two quasiphases in $\left[-\pi,\pi\right]$. For the purposes of visualization, we fold the quasiphases in the interval $\left[-\pi/2,\pi/2\right]$ to obtain the data points. In \cref{Figure_Appendix_A_MMFT_Spectrum} we reproduce \cref{Figure_XGate_Floquet_Qubits}(c) with 15 numerators to pinpoint the avoided crossing.\\

\noindent \emph{Extracting gate parameters from discrete quasiphase spectra} -- We identify the local minimum corresponding to the resonance using the following procedure:
\begin{enumerate}
    \item For a first numerator $p_0$, vary the denominator $q$ and find all discrete extrema of the quasienergy difference. \ab{For a TLS,} only one such extremum corresponds to the resonance while the others are a consequence of the folded space $\left[-\pi/2,\pi/2\right]$. We keep count of all successive triplets $\left\{\phi_{p_0}[q-1], \phi_{p_0}[q], \phi_{p_0}[q+1]\right\}$ verifying $\phi_{p_0}[q-1] \geq \phi_{p_0}[q]$ and $ \phi_{p_0}[q] \leq \phi_{p_0}[q+1]$.
    \item Given a new numerator $p_k$, we find all the triplets $\left\{\phi_{p_k}[q'-1], \phi_{p_k}[q'], \phi_{p_k}[q'+1]\right\}$ as defined above such that the interval $\left[\phi_{p_k}[q'-1],\phi_{p_k}[q'+1]\right]$ has a non-empty intersection with at least one such interval for $k-1$. 
    \item When the maximum numerator is reached, we expect to have discarded all the undesired local minima. The desired resonant frequency is located at the intersection of all ensembles:
    \begin{equation}
        \bigcap_{p_0<p_k<p_{kmax}} \left[\phi_{p_k}[q_k-1],\phi_{p_k}[q_k+1]\right].
    \end{equation}
    \end{enumerate}
The precision of the procedure is given by the range of this intersection. A simple upper  bound for this quantity is the size of the smallest segment
\begin{equation}
\left(\frac{p_{kmax}}{q_{kmax}-1}-\frac{p_{kmax}}{q_{kmax}+1}\right)\approx \frac{2p_{kmax}}{q_{kmax}^2}.
\end{equation}\\

\noindent \emph{Numerical simulation times} -- 
For the parameters of the TLS studies in \cref{Section_III_Subsection_II_Two_tone_floquet_analysis}, the ratio was estimated to be $\omega_{d1}/\omega_{d2}=p/q\approx 1/10$. To achieve a precision of up to $1$~MHz for the frequency $\omega_{d2}$ maximizing the fidelity of the $X$-Gate, a maximum numerator greater than $20$ is needed.
To reach a precision of 0.1~MHz, the maximum numerator is greater than $100$.
In that case, the propagator has to be calculated and diagonalized over 20 to 100 periods of the first drive, and for each data point. 

\begin{figure}
    \centering
    \includegraphics[width=\linewidth]{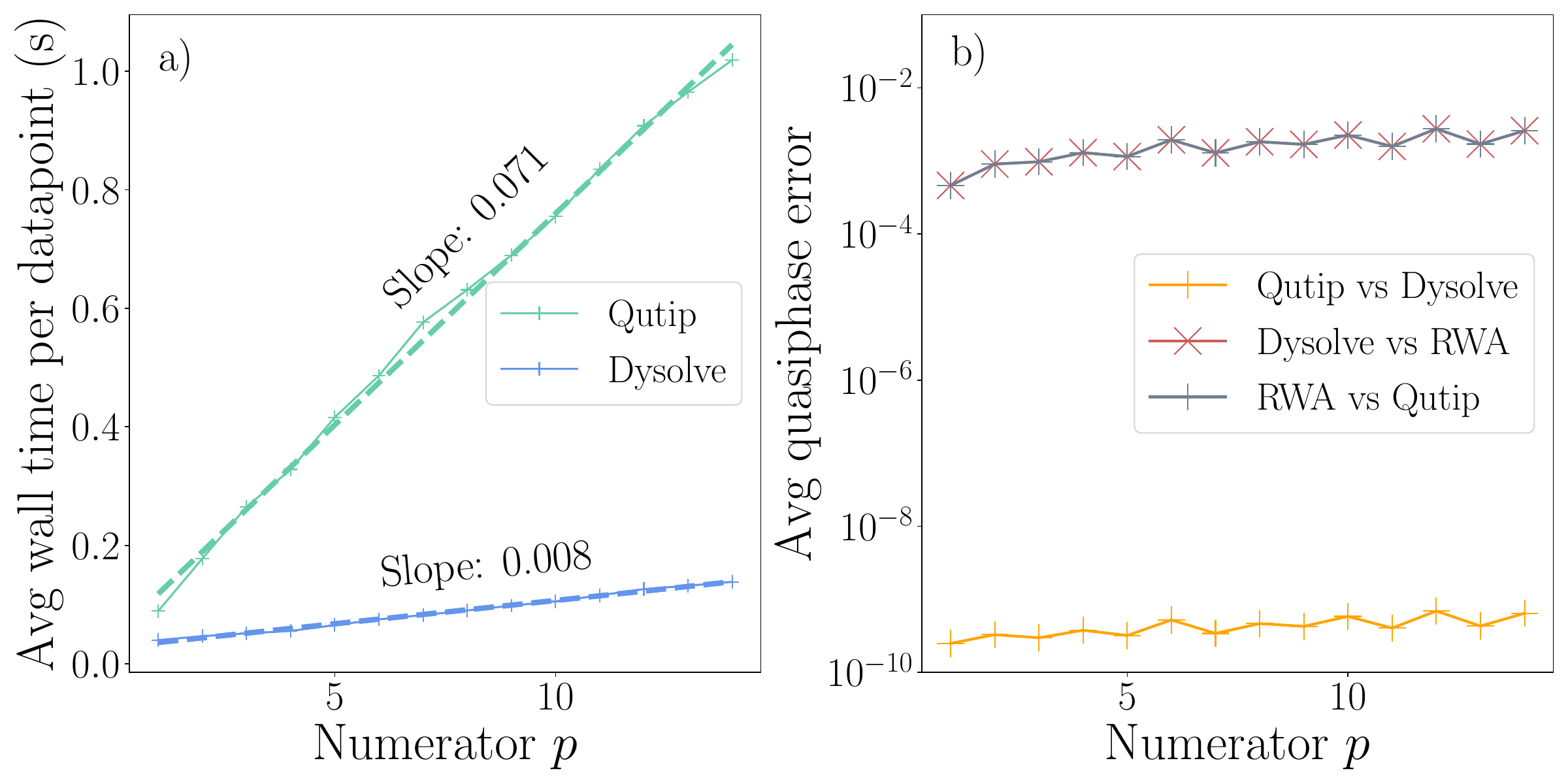}
    \caption{a) Average wall time per data point for each numerator $p$ using the QuTiP's solver (green) and Dysolve (blue). Dashed lines are linear fits to the numerical results in continuous lines. b) Comparison of the precision for the two numerical solvers and for the approximate analytical solution obtained using the RWA. The metric is the squared sum of the quasiphase difference for all data points corresponding to one numerator $p$. We find a good agreement between the two solvers (orange) and we also highlight the poor precision of the analytical result based on the rotating-wave approximation (blue and red), even for a TLS system.
    }
    \label{Figure_Appendix_A_Wall_time}
\end{figure}

With the objectives of reaching large numerators without loss of precision and of extending our study to systems with larger Hilbert space, we used the recent semi-analytic solver \textit{Dysolve} \citep{shillitoFastDifferentiableSimulation2020a} which allows for efficient numerical simulation of system dynamics in the presence of rapid oscillations. For comparison, we use QuTiP's solver for time unitary dynamics with the custom options (atol=1e-10, rtol=1e-10, nsteps=1e9) for convergence \citep{johanssonQuTiPPythonFramework2013}. 
In \cref{Figure_Appendix_A_Wall_time}, we compare the performance of Dysolve with that of QuTiP. A tenfold improvement in simulation times with equivalent or better precision is obtained with Dysolve. These gains open up a path toward precise quasiphase spectra of complex quantum systems with two drives and larger Hilbert space. 

\section{Coupler-mediated qubit-readout cavity interaction}
\label{Appendix_IV_Coupler_mediated_interaction}

To achieve the coupling Hamiltonian in \cref{TLS_Readout_Theorical_Hamiltonian} for the longitudinal readout of a Floquet qubit, we consider the circuit in \cref{Figure_Appendix_B_CQED} where a coupler-mediated qubit-cavity interaction induces 
the desired readout. We first derive the Hamiltonian of the circuit without the voltage drive on the transmon, and then add the corresponding term. 

We use a bosonic representation with a Kerr nonlinear oscillator model \cite{kochChargeinsensitiveQubitDesign2007a}. The modes $\hba$, $\hbb$, and $\hbc$ stand for readout resonator, qubit, and coupler, respectively. The three circuit elements are coupled capacitively and the Hamiltonian takes the form

\begin{align}
\label{TLS_Readout_Circuit_Hamiltonian_main}
H_\mathrm{lab}(t) &= \left( \begin{array}{ccc} \hba^\dagger & \hbb^\dagger & \hbc^\dagger \end{array} \right) 
  \left( \begin{array}{ccc}
    \Wa & g_{ab}     & g_{ca} \\
    g_{ab}    & \Wb  & g_{bc} \\
    g_{ca}    &  g_{bc}   & \Wc + \delta\Wc(t)  
    \end{array}
  \right)
  \left( 
  \begin{array}{c}
    \hba \\
    \hbb \\
    \hbc
  \end{array}
  \right) \nonumber \\
  &+\frac{\balb}{2} \hbb^{\dagger 2} \hbb^2 + \frac{\balc}{2} \hbc^{\dagger 2} \hbc^2.
\end{align}
Here, $\balb$ and $\balc$ are the qubit and coupler anharmonicities, respectively. 
The coupler is modulated parametrically with $ \delta\Wc(t)$ which is a yet-unspecified function of time, with the sole requirement that it contains no dc part: $\lim_{T\to\infty} \frac{1}{T} \int_0^{T} \delta\Wc(t) dt = 0$.

A normal-mode transformation diagonalizes the time-independent part of the quadratic Hamiltonian. This transformation amounts to putting $\hat{m} \to \sum_{l=a,b,c} u_{m l} \hat{l}$, for $m=a,b,c$. With this, the Hamiltonian consists of a diagonal quadratic form determined by the qubit, cavity, and coupler-like normal mode frequencies
\begin{equation}
\label{Eq:H0Toy}
H^{(0)} = \wa \ha^\dagger \ha + \wb \hb^\dagger \hb + \wc \hc^\dagger \hc, 
\end{equation}
and a perturbation coming from the quartic terms and the time-dependent modulation of the coupler
\begin{align}
    \lambda H^{(1)}(t) 
  = \sum_{j=b,c} \frac{\alpha_j}{2} \left(\sum_{l=a,b,c} u_{jl} \hat{l}^\dagger \right)^2 \left(\sum_{l=a,b,c} u_{jl} \hat{l} \right)^2 \nonumber \\
  +\delta\Wc(t)(\uca \ha^\dagger + \ucb \hb^\dagger + \ucc \hc^\dagger)(\uca \ha\;\, + \ucb \hb\;\, + \ucc \hc\;\,).
  \nonumber \\
\end{align}
The coupling in \cref{TLS_Readout_Theorical_Hamiltonian} can be attained by identifying $g(t) = \uca \ucb \delta \Wc(t)$, and making all other terms in $g(t)$ but $a^\dagger b + b^\dagger a$ vanish in the RWA. Moreover, one can find the resulting anharmonicities of all three modes, as well as the cross-Kerr interactions, in the RWA \cite{Petrescu2021}
\begin{align}
\alpha_j^{(1)} = \sum_{i=b,c} u_{ij}^4 \alpha_i, \; \chi^{(1)}_{jk} = \sum_{i=b,c} 2 u_{ij}^2 u_{ik}^2 \alpha_i. \label{Eq:TMCCsHI1}
\end{align}

Finally, the following term is added to the Hamiltonian to account for the yet missing qubit drive:

\begin{align}
    H_\mathrm{d}(t) = -i \epsilon_d(t) (\hbb-\hbb^\dagger) =  -i \epsilon_d(t) \sum_{l={a,b,c}} u_{bl}(\hat{l}-\hat{l}^\dagger),
\end{align}
which accounts for the control signals sent to the qubit that are off-resonantly driving the remaining two normal modes. These contributions are typically negligible.

\end{document}